\documentclass[runningheads]{llncs}
\usepackage{graphicx}
\usepackage{amssymb}
\usepackage{amsmath}
\usepackage{latexsym}
\usepackage{dsfont} 
\usepackage{xcolor}
\usepackage{hyperref}

\usepackage[colorinlistoftodos, textsize=tiny, textwidth=3.5cm]{todonotes}

\usepackage[linesnumbered,algoruled,boxed,vlined]{algorithm2e}

\SetCommentSty{mycommfont}

\newcommand{\sel}[2]{\mbox{select}\left ( {#1},{#2}\right )}
\newcommand{\join}[2]{\mbox{join}\left ({#1},~{#2} \right )}
\newcommand{\ins}[2]{\mbox{insert}({#1},{#2})}

\newcommand{\neigh}[2]{\gamma^\bullet_{{#2}}({#1})}

\newcommand{\ie}[1]{{\em i.e.}}
\newcommand{\eg}[1]{{\em e.g.}}
\newcommand{\st}[0]{\; | \;}

\newcommand{\ELIMINE}[1]{}

\newcommand{\plusequals}{\mathrel{\mathord{+}\hspace*{-1pt}\mathord{=}}}

\newcommand{\Xalgo}[1]{Algorithm~\ref{#1}}

\begin{document}
\title{Join, select, and insert: efficient out-of-core algorithms for hierarchical segmentation trees}
\titlerunning{Join, select, and insert: efficient out-of-core algorithms}
%
\author{Josselin Lefèvre \inst{1,2}, Jean Cousty \inst{1}, Benjamin Perret \inst{1}, Harold Phelippeau \inst{2}}
\authorrunning{J. Lefèvre et al.}
%
\institute{LIGM, Univ Gustave Eiffel, CNRS, ESIEE Paris, F-77454 Marne-la-Vallée, France \and
Thermo Fisher Scientific, Bordeaux, France}
\maketitle              
\begin{abstract}
Binary Partition Hierarchies (BPH) and minimum spanning trees are fundamental data structures involved in hierarchical analysis such as quasi-flat zones or watershed.
However, classical BPH construction algorithms require to have the whole data in memory, which prevent the processing of large images that cannot fit entirely in the main memory of the computer. To cope with this problem, an algebraic framework leading to a high level calculus was introduced allowing an out-of-core computation of BPHs. This calculus relies on three operations: \textit{select}, \textit{join}, and \textit{insert}. In this article, we introduce three efficient algorithms to perform these operations providing pseudo-code and complexity analysis.
\end{abstract}
\section{Introduction}

Hierarchies of partitions are versatile representations that have proven useful in many image analysis and processing problems. 
In this context, binary partition hierarchies~\cite{Salembier-Guarrido-tip2000} (BPH) built from altitude ordering and associated minimum spanning trees are key structures for several (hierarchical) segmentation methods: in particular it has been shown~\cite{cousty-et-al-ismm2013,najman-et-al-ismm2013} that such hierarchies can be used to efficiently compute quasi-flat zone (also referred as $\alpha$-trees) hierarchies~\cite{meyer1999morphological,cousty-et-al-ismm2013} and watershed hierarchies~\cite{meyer1996dynamics,cousty-et-al-ismm2013}. 
Efficient algorithms for building BPHs on standard size images are well established, but,
with the constant improvement of acquisition systems comes a dramatic increase in image resolutions, which can reach several terabytes in size.
In such case, it becomes impossible to put a single image in the main memory of a standard workstation and classical algorithms for BPHs stop working.
This creates the need for scalable algorithms to construct BPHs in an out-of-core manner to handle images that cannot fit in memory. 

In~\cite{Gazanes-Wilkinson-ijprai2019,Gotz-et-al-tpds2018,Kazemier-et-al-ismm2017}, the authors investigate distributed memory algorithms to compute min and max trees for terabytes images. In~\cite{Gigli-et-al-prl2020}, computation of minimum spanning trees of streaming images is considered. A parallel algorithm for the computation of quasi-flat zones hierarchies has been proposed in~\cite{Havel-et-al-jrtip2019}. Finally, the authors of~\cite{carlinet2022max} recently proposed massively parallel algorithms for the computation of max-trees on GPUs.
All these work rely on a common idea which is to work independently on small pieces of the space, “join” the information found on adjacent pieces, and “insert” this joint information into other pieces.

In a previous work~\cite{Cousty-et-al-DGMM2021}, the authors specifically addressed the problem of computing a BPH under the out-of-core constraint, \textit{i.e.}, when the objective is to minimize the amount of memory required by the algorithms. 
To do so, they introduced an algebraic framework formalizing the distribution of a hierarchy over a partition of the space together with three algebraic operations acting on BPHs: \emph{select}, \emph{join}, and \emph{insert}.
They showed that, when a causal partition of the space is considered, it is possible to compute the distribution of a BPH using these three operations by browsing the different regions of the partition only twice (once in a forward pass and once in a backward pass) and by requiring to have only the information about two adjacent regions in the main memory at any step of the algorithm. However, no efficient algorithm has been proposed for the three operations \emph{select}, \emph{join}, and \emph{insert}.

In this work, we propose efficient implementations for these operations. 
The proposed algorithms rely on a particular data structure to represent \emph{local} hierarchies which is designed to efficiently search and browse the nodes of the hierarchy and to store only the necessary and sufficient information required locally to compute the distribution of the BPH. 
We give algorithms, with their pseudo-code, for the three operations whose time complexity is either linear or linearithmic.  
In order to ease the presentation, we consider the particular case of 2d images, modelled as 4-adjacency graphs, but the method can be easily extended to any regular graph.
The implementation of the method in C++ and Python based on the hierarchical graph processing library Higra~\cite{perret2019higra} is available online \url{https://github.com/PerretB/Higra-distributed}.

This article is organized as follows. Section~\ref{sec:BPH} gives the definition of BPH. Section~\ref{sec:dist} recalls the notion of the distribution of a hierarchy and the calculus method that can be used to compute such distribution over a causal partition of the space. Section~\ref{sec:data} explains the proposed data structures. Section~\ref{sec:algo} presents the algorithms for the three operations \emph{select}, \emph{join}, and \emph{insert}. Finally, Section~\ref{sec:conclusion} concludes the work and gives some perspectives.

\section{Binary partition hierarchy by altitude ordering}
\label{sec:BPH}
In this section, we first remind the definitions of  hierarchy of partitions. 
Then we define the binary partition hierarchy by altitude ordering using the edge-addition operator~\cite{Cousty-et-al-DGMM2021} and we recall the bijection existing between the regions of this hierarchy and the edges of a minimum spanning tree of the graph. 

Let~$V$ be a set. A {\em partition of~$V$} is a set of
pairwise disjoint subsets of~$V$. Any element of a partition is called
a {\em region} of this partition. The {\em ground} of a
partition~$\mathbf{P}$, denoted by~$gr(\mathbf{P})$, is the union of the
regions of~$\mathbf{P}$. A partition whose ground is~$V$ is called a
{\em complete partition of~$V$}. Let~$\mathbf{P}$ and~$\mathbf{Q}$ be two
partitions of~$V$.  We say that~$\mathbf{Q}$ is a {\em refinement
  of~$\mathbf{P}$} if any region of~$\mathbf{Q}$ is included in a region
of~$\mathbf{P}$. A {\em hierarchy on~$V$} is a sequence~$(\mathbf{P}_0,
\ldots, \mathbf{P}_\ell)$ of partitions of~$V$ such that, for
any~$\lambda$ in~$\{0, \ldots, \ell - 1\}$, the
partition~$\mathbf{P}_\lambda$ is a refinement of~$\mathbf{P}_{\lambda
  +1}$. Let~$\mathcal{H} = (\mathbf{P}_0, \ldots, \mathbf{P}_\ell)$ be a
hierarchy. The integer~$\ell$ is called the {\em depth
  of~$\mathcal{H}$} and, for any~$\lambda$ in~$\{0, \ldots, \ell\}$,
the partition~$\mathbf{P}_\lambda$ is called the {\em $\lambda$-scale
  of~$\mathcal{H}$}. In the following, if~$\lambda$ is an integer
in~$\{0, \ldots, \ell\}$, we denote by~$\mathcal{H}[\lambda]$
the~$\lambda$-scale of~$\mathcal{H}$. For any~$\lambda$ in~$\{0,
\ldots, \ell\}$, any region of the~$\lambda$-scale of~$\mathcal{H}$ is
also called a {\em region of~$\mathcal{H}$}. The
hierarchy~$\mathcal{H}$ is {\em complete} if~$\mathcal{H}[0] =
\{\{x\}~|~x \in V\}$ and if~$\mathcal{H}[\ell] = \{V\}$. 
We denote by~$\mathcal{H}_\ell(V)$ the set of all hierarchies on~$V$ of
depth~$\ell$, by~$\mathcal{P}(V)$ the set of all partitions on~$V$, and
by~$2^{|V|}$ the set of all subsets of~$V$.

In the following, the symbol~$\ell$ stands for any strictly positive integer.\par
We define a graph as a pair~$G=(V,E)$ where $V$ is a finite set and $E$
is composed of unordered pairs of distinct elements in $V$. 
Each element of $V$ is called a vertex of $G$, and each element of $E$ is called an edge of $G$. 
The Binary Partition Hierarchy (BPH) by altitude ordering relies on a total order on~$E$, denoted by~$\prec$. 
Let~$k$ in~$\{1, \ldots, \ell\}$, we denote by~$u^\prec_k$ the~$k$-th element of~$E$ for the order~$\prec$. 
Let~$u$ be an edge in~$E$, the {\em rank of~$u$ for~$\prec$}, denoted by~$r^\prec(u)$, is the unique integer~$k$ such that~$u = u^\prec_k$.
We then define the {\em update of a hierarchy~$\mathcal{H}$
with respect to an edge $\{x,y\}$}, denoted by~$\mathcal{H} \oplus \{x,y\}$: with~$k$ the rank of~$\{x,y\}$,~$\mathcal{H} \oplus \{x,y\}[\lambda]$ remains 
unchanged for any~$\lambda$ in$\{0, k-1\}$ while, for any $\lambda$ in $\{k, \ldots, \ell\}$, we have~$(\mathcal{H} \oplus \{x,y\})[\lambda] = 
\mathcal{H}[\lambda] \setminus \{R_x, R_y\} \cup \{R_x \cup R_y\}$ where~$R_x$ (resp.~$R_y$) denotes the region of~$\mathcal{H}[\lambda]$ containing~$x$ (resp.~$y$).
Let~$E' \subseteq E$ and let~$\mathcal{H}$ be a hierarchy. We
set~$\mathcal{H} \boxplus E' = \mathcal{H} \oplus u_1 \oplus \ldots
\oplus u_{|E'|}$ where~$E' = \{u_1, \ldots, u_{|E'|}\}$. The binary
operation~$\boxplus$ is called the {\em edge-addition}.
Thanks to this operation, we can define formally the BPH for~$\prec$.
Let~$X$ be a set, we denote by~$\perp_X$ the hierarchy defined by
$\perp_X[\lambda] = \{\{x\} \st x \in X\}$, for any~$ \lambda$
in~$\{0, \ldots \ell\}$. The BPH for~$\prec$, denoted by~$\mathcal{B}^\prec$ is the hierarchy~$\perp_X \boxplus E$.\par

\begin{figure}[htb]
    \begin{center}
      \begin{tabular*}{\linewidth}{@{\extracolsep{\fill}}c c c}
        \includegraphics[width=0.20\linewidth]{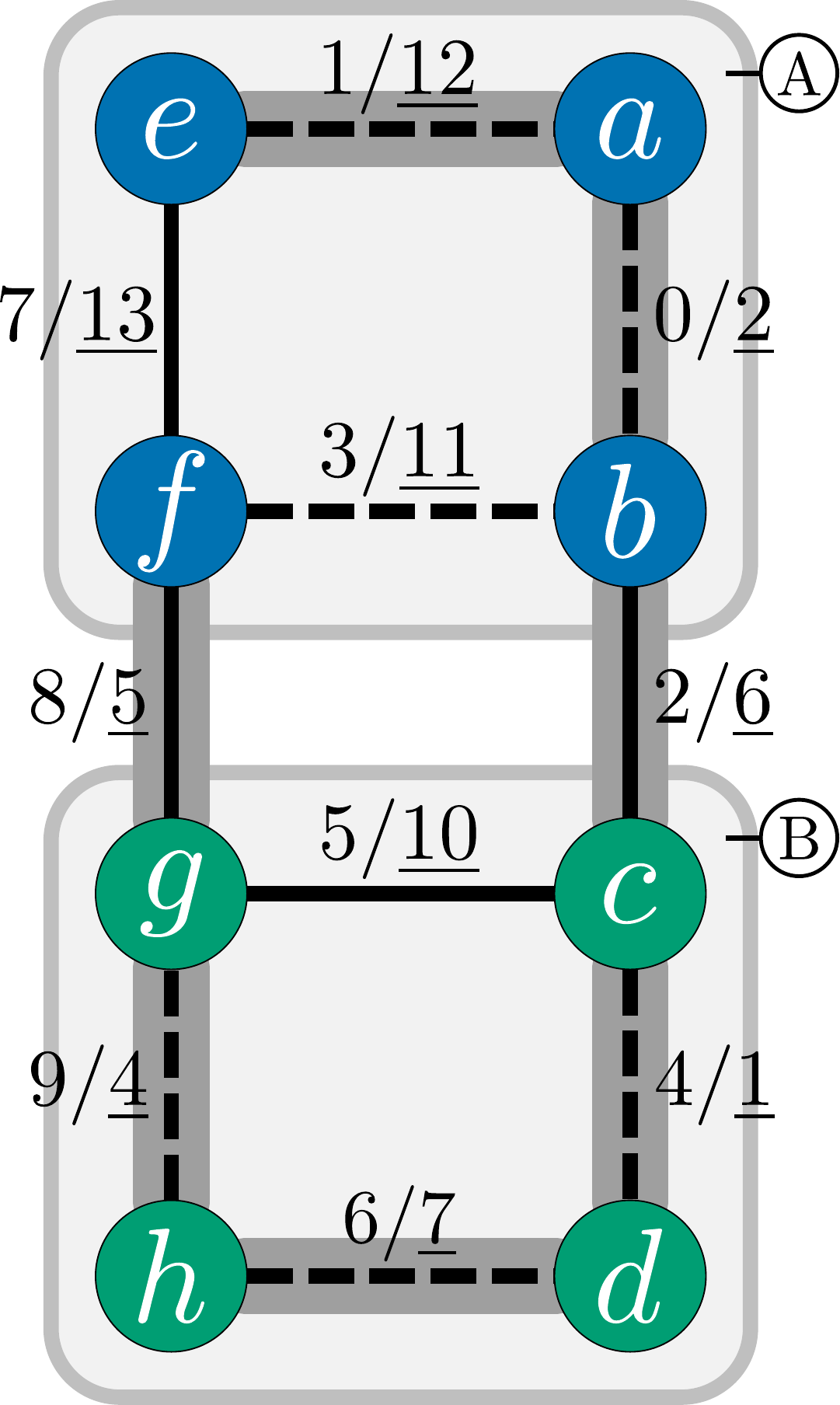}&
        \includegraphics[width=0.35\linewidth]{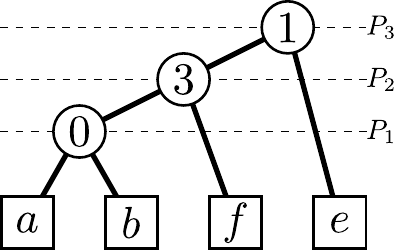}&
        \includegraphics[width=0.35\linewidth]{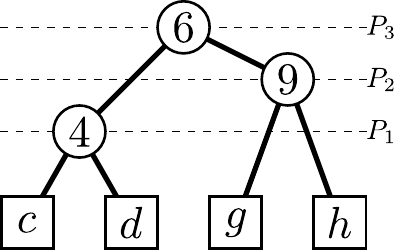}\\
      $G$ & $\mathcal{X}$ & $\mathcal{Y}$
\end{tabular*}
\caption{\label{fig:bpt}$G$ a 4-adjacency graph divided into two slices~$A$ and~$B$ (respectively blue and green). Each edge of~$G$ is associated with a pair (\textit{index, \underline{weight}}). The two slices are separated by their common neighborhood \ie{} edges 8 and 2.  Hierarchy~$\mathcal{X}$ (respectively~$\mathcal{Y}$) is the BPH built on~$A$ (respectively~$B$). Indices associated with non-leaf nodes of the BPHs correspond to the indices of their corresponding building edges represented by dashed edges in~$G$. We can note that MSTs built on slices (dashed edges) are not sub-trees of the complete MST (shown as shadow). In consequence, edge 3 is part of the hierarchy~$\mathcal{X}$ when it should not be.}
 \end{center}
\vspace{-2em}
\end{figure}

Let~$\mathcal{B}^\prec$ be a binary partition by altitude ordering,~$\mathcal{R}$ be a region of~$\mathcal{B}^\prec$ and~$\mathcal{R}^\star$ be the set of non-leaf regions of~$\mathcal{B}^\prec$. The \textit{rank} of~$\mathcal{R}$, denoted by~$r(\mathcal{R})$, is the lowest integer~$\lambda$ such that~$\mathcal{R}$ is a region of~$\mathcal{B}^\prec[\lambda]$.
We consider the map~$\mu$ from~$\mathcal{R}^\star$ in~$E$ such that, for any non-leaf region~$\mathcal{R}$ of~$\mathcal{B}^\prec$, we have~$\mu^\prec(\mathcal{R}) = u^\prec_{r(\mathcal{R})}$. We say
that~$\mu^\prec(\mathcal{R})$ is the \textit{building edge} of~$\mathcal{R}$. Building edges of the binary partitions hierarchy defines a minimum spanning tree of an edge-weighted graph. In Figure~\ref{fig:bpt},~$\mathcal{Y}$ is the BPH built on the 4-adjacency graph~$B$. Non-leaf nodes of~$\mathcal{Y}$ correspond to the edges of the minimum spanning tree of~$B$ (dashed edges).

\section{Distributed hierarchies of partitions on causal partition}
\label{sec:dist}

In this section, we recall the definition of the distribution of a BPH on a sliced graph and   the principle of its calculus in an out-of-core manner. 
Intuitively, distributing a hierarchy consists in splitting it into a set of smaller trees such that: 1) each smaller tree corresponds to a \textit{selection} of a sub part of whole tree that intersects a slice of the graph and 2) the initial hierarchy can be reconstructed by ``gluing" those smaller trees. 

Let~$V$ be a set. The operation {\em sel} is the map from~$2^{|V|}
\times \mathcal{P}(V)$ to~$\mathcal{P}(V)$ which associates to any
subset~$X$ of~$V$ and to any partition~$\mathbf{P}$ of~$V$ the
subset~$\mbox{sel}(X,\mathbf{P})$ of~$\mathbf{P}$ which contains every
region of~$\mathbf{P}$ that contains an element of~$X$. The operation
{\em select} is the map from~$2^{|V|} \times \mathcal{H}_\ell(V)$
in~$\mathcal{H}_\ell(V)$ which associates to any subset~$X$ of~$V$ and
to any hierarchy~$\mathcal{H}$ on~$V$ the
hierarchy~$\sel{X}{\mathcal{H}} =
(\mbox{sel}(X,\mathcal{H}[0]), \ldots,
\mbox{sel}(X,\mathcal{H}[\ell]))$.\par

We are then able to define the distribution of a hierarchy thanks to \textit{select}.
Let~$V$ a set, let~$\mathbf{P}$ be \textit{a complete partition on}~$V$ and 
let~$\mathcal{H}$ be a hierarchy on~$V$. The distribution of~$\mathcal{H}$ 
over~$\mathbf{P}$ is the set~$\{\sel{R}{\mathcal{H}}~|~R \in \mathbf{P}\}$ and for any region~$R$ of~$\mathbf{P}$,~$\sel{R}{\mathcal{H}}$ is called a \textit{local hierarchy (of~$\mathcal{H}$ on $R$)}. 

The calculus introduced in~\cite{Cousty-et-al-DGMM2021} aims to compute the distribution of a BPH over a partition of the space.
In this article, we consider the special case of a 4 adjacency graph representing a 2d image that can be divided into slices, and we are interested in computing a distribution of the BPH over those slices.
It should be noted that this is not a limiting factor, and the method can easily be adapted to any regular grid graph. 

Let~$h$ and~$w$ be two integers representing the height and the width of an image. In the following, the set~$V$ is the Cartesian product of~$\{0, \cdots, h-1\} \times \{0, \cdots, w-1\}$. 
Thus, any element~$x$ of~$V$ is a pair~$x=(x_i, x_j)$ such that~$x_i$ and~$x_j$ are the coordinates of~$x$. 
In the 4-adjacency grid, the set of all edges $E$ is equal to $\{\{x,y\} \in V \st \lvert x_i - y_i \rvert + \lvert x_j - y_j\rvert \leq 1\}$.
Let $k$ be a positive integer, the causal partition of~$V$ is the sequence~$(S_0, \ldots, S_k)$ such that for any~$t$ in~$\{0, \cdots, k\}$,~$S_t = \{(i,j)\in V \st t\times \frac{w}{k} \leq i < (t+1)\times \frac{w}{k}\}$.
Each element of this partition is called a \emph{slice}. The set of vertices at the interface between two neighbor slices~$A$ and~$B$ and belonging to~$A$ is noted~$\neigh{A}{B}$.
The major advantage of considering this partition over a regular graph is that each subset of~$V$ or~$E$ can be computed on the fly efficiently from a computational and memory point of view. 
 
 \begin{algorithm}[htb]
  \DontPrintSemicolon
  \KwData{A graph~$(V,E)$, a total order~$\prec$ on~$E$, and a causal
    partition~$(S_0, \ldots, S_k)$ of~$V$\; }
  \KwResult{$\{\mathcal{B}^\downarrow_0, \ldots,
    \mathcal{B}^\downarrow_k\}$: the distribution of the BPH~$\mathcal{B}^\prec_V$ over~$\{S_0, \ldots,
    S_k\}$.}

 $\mathcal{B}^\uparrow_0$ := $\mathcal{B}^\prec_{S_0}$
 \tcp*{call PlayingWithKruskal algorithm}

 \ForEach( \tcp*[f]{Causal traversal of the slices}){i \textbf{from} $1$ \textbf{to} $k$}{ 
   Call PlayingWithKruskal algorithm to compute $\mathcal{B}^\prec_{S_i}$\; \label{line:pwk_bprec}
 $\mathcal{M}^\uparrow_i$ :=
   $\join{\sel{\neigh{S_{i-1}}{S_i}}{\mathcal{B}^\uparrow_{i-1}}}{\sel{\neigh{S_i}{S_{i-1}}}{\mathcal{B}^\prec_{S_i}}}$\;
   $\mathcal{B}^\uparrow_i$ := $\ins{\sel{\neigh{S_i}{S_{i-1}}}{\mathcal{M}^\uparrow_i}}{\mathcal{B}^\prec_{S_i}}$
 }

  $\mathcal{B}^\downarrow_k$ := $\mathcal{B}^\uparrow_k $; $\mathcal{M}^\downarrow_k$ :=
    $\mathcal{M}^\uparrow_k $
 
 \ForEach( \tcp*[f]{Anticausal traversal of the slices}){i \textbf{from} $k-1$ \textbf{to} 0 }{ 
   $\mathcal{B}^\downarrow_i$ := $\ins{\sel{\neigh{S_i}{S_{i+1}}}{\mathcal{M}^\downarrow_{i+1}}}{\mathcal{B}^\uparrow_{i}}$\;
   \lIf{$i > 0$}{$\mathcal{M}^\downarrow_i$ := $\ins{\sel{\neigh{S_i}{S_{i-1}}}{\mathcal{B}^\downarrow_{i}}}{\mathcal{M}^\uparrow_i}$ }
 }
\caption{\label{algo:bptooc} Out-of-core binary partition hierarchy~\cite{Cousty-et-al-DGMM2021}.}
\end{algorithm}

Given this causal partition, \Xalgo{algo:bptooc} allows computing the local hierarchies of the BPH of the complete graph on each slice. This algorithm can be divided in two parts: causal and anti-causal traversal of the slices. 
Each of these parts relies on the same idea. First, start with the causal traversal. 
Given a causal partition of~$V$ into~$k+1$ slices, for any~$i$ in~$\{1, \cdots, k\}$ compute the BPH on~$S_i$ with a  call to the algorithm presented in~\cite{najman-et-al-ismm2013} hereafter called PlayingWithKruskal (line~\ref{line:pwk_bprec}).
Then, select the part of this hierarchy containing the vertices adjacent to the previous slice and join it with the part of the hierarchy associated to the previous slice containing the vertices adjacent to the current slice, leading to the ``merged" hierarchy denoted by~$\mathcal{M}^\uparrow_i$ (line 4). The merged hierarchy is then inserted in the BPH which gives~$\mathcal{B}^\uparrow_i$ (line 5).
The hierarchies~$\mathcal{B}^\uparrow_i$ associated to slice~$i$ misses the information located in slices of higher indices, and consequently only the last local hierarchy~$\mathcal{B}^\uparrow_k$ is correct \ie{}~$\mathcal{B}^\uparrow_k = \sel{S_k}{\mathcal{H}}$.
In order to compute the valid distribution, and after having spread information in the causal direction, information must be back propagated in the reverse anti-causal direction so that each local hierarchy is enriched with the global context (lines 7 to 9).

\section{Data structures}
\label{sec:data}

In this section, we  present the data structures used in the 
algorithms defined in the following sections.
These data structures are designed to contain only the necessary and sufficient information so that we never need to have all the data in the main memory at once. 
The data structure representing a local hierarchy assumes that the nodes of the hierarchy are indexed in a particular order and relies on three ``attributes": 1) a mapping of the indices from the local context (a given slice) to the global one (the whole graph) noted~$\mathcal{H}$\texttt{.map}, 2) the parent array denoted by~$\mathcal{H}$\texttt{.par} encoding the parent relation between the tree nodes, and 3) an array~$\mathcal{H}$\texttt{.weights} giving, for each non-leaf-node of the tree, the weight of its corresponding building edge. 

More precisely, given a  binary partition hierarchy~$\mathcal{H}$ with~$n$ regions, every integer between $0$ and~$n-1$ is associated to a unique region of~$\mathcal{H}$.  Moreover, this indexing of the regions of~$\mathcal{H}$ follows a topological order such that: 
1) any leaf region is indexed before any non-leaf region; 
2) two leaf regions~$\{x\}$ and~$\{y\}$ are sorted with respect to an arbitrary order on the element~$V$, called the \textit{raster scan order} of~$V$. Thus~$\{x\}$ has an index lower than~$\{y\}$ if~$x$ is before~$y$ with respect to the raster scan order; 
and 3) two non-leaf regions are sorted according to their rank, \emph{i.e.}, the order of their building edges for~$\prec$.
This order can be seen as an extension of the order~$\prec$ on~$E$ to the set~$V \cup E$ that enables 1) to efficiently browse the nodes of a hierarchy according to their scale of appearance in the hierarchy and 2) to efficiently match regions of $V$ with the leaves of the hierarchy. By abuse of notation, this extended order is also denoted by~$\prec$ in the following. 

To keep track of the global context, a link between the indices in the local tree and the global indices in the whole graph is stored in the form of an array \texttt{map} which associates: 
1) to the index~$i$ of any leaf region~$R$, the vertex~$x$ of the graph~$G$ such that~$R=\{x\}$, \ie{} \texttt{map[i]=x}; and
2) to the index~$i$ of any non-leaf region~$R$, its building edge, \ie{} \texttt{map[i]=$\mu^\prec(R)$}.

The parent relation of the hierarchy is stored thanks to an array \texttt{par} such that \texttt{par[i]=j} if the region of index~$j$ is the parent of the region of index~$i$. 

The binary partition hierarchy is built for a particular ordering~$\prec$ of the edges of~$G$. In practice, this ordering is induced by weights computed over the edges of~$G$. 
To this end, we store an array \texttt{weights} of~$\lvert \mathcal{R}^\star(\mathcal{H)} \rvert$, \ie{} the number of non-leaf regions, elements such that, for every region~$R$ in~$\mathcal{R}^\star(\mathcal{H})$ of index~$i$, \texttt{weights[i]} is the weight of the building edge $\mu^\prec(R)$ of region~$R$. The edges can then be compared according to the following total order induced by the weights: we set $u \prec v$ if the weight of~$u$ is less than the one of~$v$ or if $u$ and $v$ have equal weights but $u$ comes before $v$ with respect to the raster scan order.

\section{Algorithms}
\label{sec:algo}

\textbf{Select.} In this part, we give an algorithm to compute the result of the \textit{select} operation. 
This operation consists in ``selecting" the part of a given hierarchy intersecting a subset of the space. In \Xalgo{algo:bptooc}, \textit{select} takes as input a set of vertices located at the ``border" of a slice and a hierarchy in order to obtain a smaller ``border hierarchy". 

Select algorithm proceeds in 3 steps:

\begin{enumerate}
    \item {\em Lines \ref{line:markleavesin}-\ref{line:markleavesout}.} Mark any leaf-node of~$\mathcal{H}$ that corresponds to an element of~$X$, \ie{} any leaf-region~$\{x\}$ with~$x \in X$;
    \item {\em Lines \ref{line:markin}-\ref{line:markout}.} Traverse the hierarchy from leaves to root and mark any node that is a parent of a marked node;
    \item {\em Lines \ref{line:buildin}-\ref{line:buildout}.} Build the hierarchy~$\mathcal{S}$ whose nodes are only marked nodes of~$\mathcal{H}$.
\end{enumerate}

\begin{algorithm}
\DontPrintSemicolon
\KwData{$\mathcal{H}$: a hierarchy, $X$: set of selected nodes st.~$X \subset gr(\mathcal{H})$}
\KwResult{$\mathcal{S}$: the hierarchy $\sel{X}{\mathcal{H}}$}
Initialize an array \texttt{mark} to false for every region of~$\mathcal{H}$\;
$i := 0; j := 0$ \tcp*[f]{$i$ iterates over~$\mathcal{X}$ and~$j$ over the leaves of~$\mathcal{H}$}\;
\While{$i < \lvert X \rvert$ \textbf{and} $\ j < \lvert \mathcal{H}.leaves \rvert$}{ \label{line:markleavesin}
    \If{$X[i] = \mathcal{H}.map[j]$}{
        $mark[j] := true$\;
        $i := i+1$\;
    }
    $j := j+1$\;
} \label{line:markleavesout}
\ForEach{$n$ \textbf{from} 0 \textbf{to} $\lvert \mathcal{H} \rvert -1$}{\label{line:markin}
    \lIf{$mark[n] \, = \, true$}{ $mark[\mathcal{H}$.par$[n]]:=true$}\label{line:markout}
}
$n_\mathcal{S} := 0$\;
\ForEach{$n$ \textbf{from} 0 \textbf{to} $\lvert \mathcal{H} \rvert -1$}{\label{line:buildin}
    \If{$mark[n] \, = \, true$}{ 
        $\mathcal{S}$.par[$n_\mathcal{S}] := \mathcal{H}$.par[n]\;
        $\mathcal{S}$.map[$n_\mathcal{S}] := \mathcal{H}$.map[n]\;    
        \If{$n \in \mathcal{R}^\star(\mathcal{H})$}{
            $\mathcal{S}$.weight[$n_\mathcal{S} - \lvert X \rvert] := \mathcal{H}$.weight[$n_\mathcal{S} - \lvert \mathcal{H}.leaves \rvert$]
        }
    }
    $n_\mathcal{S} := n_\mathcal{S} + 1$\label{line:buildout}
}
\Return{$\mathcal{S}$}\;
\caption{{\sc select}}
\label{algo:select}
\end{algorithm}

In \Xalgo{algo:select} we assume that~$X$ is sorted and that~$X \subset gr(\mathcal{H})$, which is always the case in \Xalgo{algo:bptooc}. For each element $X[i]$ of $X$, we search for the index~$j$ of a leaf of~$\mathcal{H}$ mapped to~$X[i]$, \ie{} such that~$\mathcal{H}.\texttt{map[j]} = X[i]$. To this end, it is necessary to make a traversal of the leaves of~$\mathcal{H}$. As mentioned before, the leaves correspond to the first indices by construction.
The first step can then be performed in linear time with respect to the number of leaf-regions of~$\mathcal{H}$. The second step consist in traversing the whole hierarchy from leaves to root in order to mark every region of~$\mathcal{H}$ which belongs to~$\sel{X}{\mathcal{H}}$ \ie{} regions parent of a marked one. The complexity of this step is therefore linear with respect to the number of regions of~$\mathcal{H}$. Finally, the last step boils down to extracting the hierarchy~$\sel{X}{\mathcal{H}}$ from the marked nodes. For this a new hierarchy is created by traversing~$\mathcal{H}$ again. As the traversal is done by increasing order of index, the properties relating to the weights of the building edges and order of appearance of regions are preserved. The complexity of this last step is linear with respect to the number of regions of~$\mathcal{H}$. Thus, \Xalgo{algo:select} has a linear~$O(n)$ complexity, where~$n$ is the number of regions of~$\mathcal{H}$.\par

\textbf{Join.} Formally the {\em join of~$\mathcal{X}$ and~$\mathcal{Y}$}, denoted
by~$join(\mathcal{X}, \mathcal{Y})$, is the hierarchy defined
by~$join(\mathcal{X}, \mathcal{Y}) = (\mathcal{X} \sqcup \mathcal{Y})
\boxplus F$, where~$F$ is the common neighborhood of the grounds
of~$\mathcal{X}$ and of~$\mathcal{Y}$, and~$\sqcup$ denotes the supremum on hierarchies (see~\cite{Ronse-jmiv2008}). 
Intuitively, this operation merges two hierarchies according to their common neighborhood, that is the set of edges linking their grounds.
In~\cite{Havel-et-al-jrtip2019}, the authors proposed an algorithm that can be used to successively add edges of the common neighborhood. Intuitively, to add an edge, the hierarchy is updated while climbing the branches associated with the edge extremities. The worst-case complexity is then linear with respect to the size of the hierarchy for adding a single edge. Thus the overall complexity of such join procedure would be~$O(k\times n)$ where~$n$ is the size of the hierarchies and~$k$ is the number of edges in the common neighborhood. In this section, we drop the multiplicative dependency in the size of the neighborhood at the cost of introducing a sorting of~$F$ and we present an algorithm whose complexity is quasi-linear with respect to the size~$n$ of the hierarchies and linearithmic with respect to the number~$k$ of edges in~$F$.

\begin{algorithm}[htb]
\DontPrintSemicolon
\KwData{$\mathcal{X}$$ and~$$\mathcal{Y}$: two hierarchies, $F$ common neighborhood of $gr(\mathcal{X})$ and $gr(\mathcal{Y})$.}
\KwResult{A collection $Q_D = \join{\mathcal{X}}{\mathcal{Y}}$}
\lForEach{$node \ n_i \ of \ \mathcal{X}$}{$Q_D$.MakeSet$(i)$} \label{line:dsin}
\lForEach{$node \ n_i \ of \ \mathcal{Y}$}{$Q_D$.MakeSet$(i+\lvert \mathcal{X}$.leaves$  \rvert)$} \label{line:dsout}
aDescendent$(\mathcal{X},0)$\; \label{line:arootin}
aDescendent$(\mathcal{Y},\lvert \mathcal{X}.leaves \rvert)$\; \label{line:arootout}
$F := $sort$ (F)$\; \label{line:sort}
$i_1 := \lvert \mathcal{X}.leaves \rvert; \ i_2 :=\lvert \mathcal{Y}.leaves \rvert; \ i_3 := 0$\;
\While{$i_1 < \lvert \mathcal{X} \rvert$ \textbf{or}~$i_2 < \lvert \mathcal{Y} \rvert$ \textbf{or}~$i_3 < \lvert F \rvert$}{ \label{line:pwkin}
    \If{$F[i_3] \prec \mathcal{X}$.map$[i_1]$ \textbf{and}~$F[i_3] \prec \mathcal{Y}$.map$[i_2]$}{
        $(x,y) := F[i_3]; \ m := F[i_3];\ w:= weight(F[i_3]);\ i_3 \plusequals 1$\;
    }
    \ElseIf{$\mathcal{X}$.map$[i_1] \prec \mathcal{Y}$.map$[i_2]$}{
        $(x,y) := \mathcal{X}$.desc$[i_1]; \ m := \mathcal{X}$.map$[i_1];\ w := \mathcal{X}$.weight$[i_1];\ i_1 \plusequals 1$
    }
    \Else{
        $(x,y) := \mathcal{Y}$.desc$[i_2]; \ m := \mathcal{Y}$.map$[i_2];\ w := \mathcal{Y}$.weight$[i_2]; \ i_2 \plusequals 1$ \label{line:joinEndBrowsing}
    }
    $c_x := Q_D$.FindCanonical$(x); \ c_y := Q_D$.FindCanonical$(y)$\;
    \If{$c_x != c_y$\label{line:joinPlayingBegin}}{
        $n := Q_D$.Union$(c_x,c_y)$;\
        $Q_D$.map$[n] := m$\;
        $Q_D$.weight$[n-(\lvert \mathcal{X}.leaves \rvert + \lvert \mathcal{Y}.leaves \rvert)] := w$ \label{line:pwkout}
    }
}

\caption{{\sc Join}}
\label{algo:join}
\end{algorithm}

\begin{figure}[htb]
    \begin{center}
      \begin{tabular*}{1\linewidth}{@{\extracolsep{\fill}}c c c c}
        \includegraphics[width=0.135\linewidth]{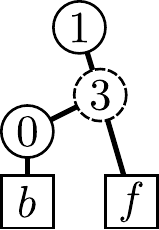}&
        \includegraphics[width=0.135\linewidth]{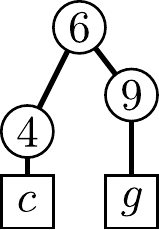}&
        \includegraphics[width=0.31\linewidth]{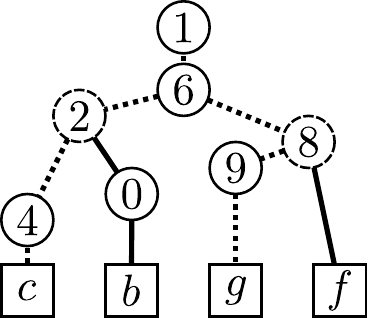}&
        \includegraphics[width=.31\linewidth]{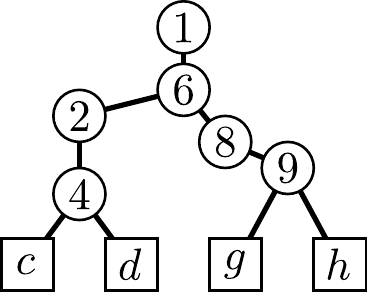}\\
      $\mathcal{X}^\prime$ & $\mathcal{Y}^\prime$ & $\mathcal{J}=\join{\mathcal{X}^\prime}{\mathcal{Y}^\prime}$ & $\mathcal{I} = \ins{\mathcal{J}^\prime}{\mathcal{Y}}$
\end{tabular*}
\caption{\label{fig:join}The hierarchy~$\mathcal{J}$ is build by computing the join over $\mathcal{X}^\prime=\sel{\{c, h\}}{\mathcal{X}}$ and~$\mathcal{Y}^\prime=\sel{\{d, i\}}{\mathcal{Y}}$ (border trees computed from BPHs of Figure~\ref{fig:bpt}). We can see that the node 3 of~$\mathcal{X}^\prime$ not longer appear in the joint tree in the favor of nodes corresponding to the common neighborhood of the grounds of~$\mathcal{X}$ and~$\mathcal{Y}$ \ie{} nodes 2 and 8. That is to say, that by taking into account the topological order on the edges of the MSTs associated with the border trees and the common neighborhood, 3 does not belongs to the BPH. $\mathcal{I}$ is then build by inserting the hierarchy~$\mathcal{J}^\prime = \sel{\{c, g\}}{\mathcal{J}}$ into the hierarchy~$\mathcal{Y}$. Given the Definition 16 of~\ref{algo:bptooc},~$\mathcal{I}$ is a ``correct" local hierarchy for the tile B.}
 \end{center}
\end{figure}

A detailed presentation of the proposed algorithm is given in \Xalgo{algo:join} which calls auxiliary functions presented in \Xalgo{algo:qd}. Intuitively, in order to compute the join of two hierarchies~$\mathcal{X}$ and~$\mathcal{Y}$, \Xalgo{algo:join} consists in ``emulating" PlayingWithKruskal algorithm on the graph obtained from (i) the edges associated to the non-leaf nodes of~$\mathcal{X}$ and of~$\mathcal{Y}$ and (ii) the edges in the common neighborhood~$F$ of~$\mathcal{X}$ and~$\mathcal{Y}$. Therefore, all these edges are considered in increasing order with respect to~$\prec$ and, for each edge, it is decided if this edge must be considered or not in the creation process of the join hierarchy. The decision is made based on the potential creation of a cycle if this edge were added during the minimum spanning tree creation process. We can thus see on the Figure~\ref{fig:join} that the node 3 has been added to~$\mathcal{X}$ by construction but that it is then discarded during the construction of the joined hierarchy. Potential-cycles creation is efficiently checked with Tarjan Union-Find data structures as in Kruskal's algorithm. A main observation can be made to highlight the difference between the situation encountered in the contexts of join algorithm  and PlayingWithKruskal algorithms: in the context of join, some edges, which are associated to the nodes of the hierarchies~$\mathcal{X}$ and~$\mathcal{Y}$, are made of vertices that do not belong to the underlying space ({\em i.e.}, the common neighborhood of the slices supporting the grounds of $\mathcal{X}$ and~$\mathcal{Y}$).  When such edge is found, the standard algorithm can be shortcut leading to a modified version of the PlayingWithKruskal auxiliary functions presented in \Xalgo{algo:qd}. Compared to original functions, the only change is the insertion of the if test at line \ref{fun:unionIf}. This test detects the edges for which a shortcut must occur based on an attribute called \texttt{desc}. This attribute is pre-computed for every node of~$\mathcal{X}$ and of~$\mathcal{Y}$ by the auxiliary function \texttt{aDescendent}. Overall, the following steps are performed in \Xalgo{algo:join}:
\begin{itemize}
    \item {\em Lines \ref{line:dsin}-\ref{line:dsout}.} Initialize the Union-find data structures;
    \item {\em Lines \ref{line:arootin}-\ref{line:arootout}.} Compute the attribute \texttt{desc} for both~$\mathcal{X}$ and~$\mathcal{Y}$;
    \item  {\em Lines \ref{line:sort} to \ref{line:joinEndBrowsing}.} Browse the edges in increasing order. Observe that it implies sorting the edges in~the common neighborhood $F$ of~$\mathcal{X}$ and~$\mathcal{Y}$ in increasing order for~$\prec$ (non-leaf-nodes of~$\mathcal{X}$ and~$\mathcal{Y}$ are already sorted by construction);
    \item {\em Lines \ref{line:joinPlayingBegin}-\ref{line:pwkout}} Apply  PlayingWithKruskal steps, calling the modified version of the auxiliary functions.
\end{itemize}

\begin{algorithm}[ht]
    \SetAlgoLined\DontPrintSemicolon

    \SetKwFunction{func}{func}\SetKwFunction{proc}{proc}
\tcp{The functions called hereafter on $Q_{T}$ and $Q_{BT}$ are those described in \cite{najman-et-al-ismm2013} }

    \SetKwProg{makeset}{Procedure}{}{}
    \makeset{$Q_D.MakeSet(q)$}{
        $Q_D.Root[q]:=q; \ Q_{BT}.MakeSet(q); \ Q_{T}.MakeSet(q)\;$}\

    \SetKwProg{findcanonical}{Function}{}{}
    \findcanonical{$Q_D.FindCanonical(q)$}{
        \KwRet $Q_T.FindCanonical(q)$}\
    
    \SetKwProg{union}{Function}{}{}
    \union{$Q_D.Union(c_x,c_y) $}{
        $t_u := Q_D.Root[c_x]$; \ $Q_{BT}.par[t_u] := Q_BT.size$\;
        \lIf{$c_y = -1$\label{fun:unionIf}}{
            $Q_D.Root[c_x] := Q_{BT}.size$
        }
        \Else{\label{fun:unionelse} 
            $t_v := Q_D.Root[c_y]$;\ 
            $Q_{BT}.par[t_v] := Q_D.size$\;
            $c := Q_{T}.Union(c_x,c_y)$;\
            $Q_D.Root[c] := Q_{BT}.size$\; }
        $Q_{BT}.MakeSet(Q_{BT}.size)$\;
        \KwRet $Q_{D}.size-1$}\;
    \SetKwProg{aDescendent}{Function}{}{}
    \aDescendent{aDescendent($\mathcal{H}$: a hierarchy, $s$: shift)}{
         \tcp{set the attribute $\mathcal{H}$.\texttt{desc}: an array that maps to each node $n$ of~$\mathcal{H}$ two leaves that are descendants of the two children of $n$.}
         \lForEach{$node \ n \ of \ \mathcal{H}$}{
            $\mathcal{H}$.desc[n]$ := ( -1,-1 )$
        }
        \lForEach{$leaf \ node \ n \ of \ \mathcal{H}$}{
            $\mathcal{H}$.desc[n].first$ := n + s$
        }
        \ForEach{non-root non-leaf node  $n$  of $\mathcal{H}$ in increasing order for~$\prec$}{
            $p := \mathcal{H}.par[n]$\;
            \lIf{$\mathcal{H}$.desc[p].first$ = -1$}{$\mathcal{H}$.desc[p].first$ := \mathcal{H}$.desc[n].first}
            \lElse{$\mathcal{H}$.desc[p].second$ := \mathcal{H}$.desc[n].first}
        }
    }
    
\caption{{\sc Auxiliary functions for \textit{join} algorithm}}
\label{algo:qd}
\end{algorithm} 

The first step complexity is linear with respect to the number of elements of $gr(\mathcal{X}) \cup gr(\mathcal{Y})$. The second step uses the auxiliary function \texttt{aDescendent} to compute attributes \texttt{desc} for both~$\mathcal{X}$ and~$\mathcal{Y}$. It should be noted that this last function takes a parameter \textit{shift} which allows to index the leaves of the second hierarchy after those of the first. For each node of the two hierarchies, an attribute is computed during a leaves to root traversal which gives a linear complexity with respect to the number of regions of each hierarchy. Third step requires to sort the edges of~$F$ with respect to~$\prec$ before browsing the edges which implies a complexity of~$O(k \times log(k) + \lvert \mathcal{X} \rvert + \lvert \mathcal{Y} \rvert)$ with~$k$ the number of edges in~$F$.
The fourth step is equivalent to PlayingWithKruskal algorithm in terms of complexity. Then, its complexity is~$O(m \times \alpha(n))$ where~$m$ is sum of the number of edges in~$F$ and the number of non-leaf nodes of~$\mathcal{X}$ and~$\mathcal{Y}$, where~$n$ is the number of leaf nodes in~$\mathcal{X}$ and~$\mathcal{Y}$ and where~$\alpha()$ is the inverse Ackermann function which grows sub-logarithmically.\par 

\textbf{Insert.} In this part, we present an algorithm to compute the hierarchy~$\mathcal{Z} = \ins{\mathcal{X}}{\mathcal{Y}}$. We assume that~$\mathcal{X}$ is \textit{insertable} in~$\mathcal{Y}$ \ie{} for any~$\lambda$ in~$\{0, \ldots, \ell\}$, for any region~$Y$
  of~$\mathcal{Y}[\lambda]$,~$Y$ is either included in a region
  of~$\mathcal{X}[\lambda]$ or is included in~$V \setminus
  gr(\mathcal{X}[\lambda])$. This assumption holds true at each call to \textit{insert} in \Xalgo{algo:bptooc}. The {\em insertion
    of~$\mathcal{X}$ into~$\mathcal{Y}$} is the
  hierarchy~$\mathcal{Z}$, such that, for any~$\lambda$ in~$\{0,
  \ldots, \ell\}$,~$\mathcal{Z}[\lambda] = \mathcal{X}[\lambda] \cup
  \{R \in \mathcal{Y}[\lambda] \st R \cap gr(\mathcal{X}[\lambda]) =
  \emptyset \}$. Algorithm~\ref{algo:insert}, presented hereafter, computes the insertion~$\mathcal{Z}$ of~$\mathcal{X}$ into~$\mathcal{Y}$. From a high level point of view, it proceeds in two main steps:
  
\begin{algorithm}[H]
\DontPrintSemicolon
\KwData{$\mathcal{X}$ and~$\mathcal{Y}$: two hierarchies such that~$\mathcal{X}$ insertable in~$\mathcal{Y}$.}
\KwResult{$\mathcal{Z}$: the hierarchy~$\ins{\mathcal{X}}{\mathcal{Y}}$}
$x:=0; \ y:=0;\ z:=0$; \tcp{indices for the nodes/regions of~$\mathcal{X}$,~$\mathcal{Y}$, and~$\mathcal{Z}$}
Initialize an array \textit{InZ} of~$|\mathcal{Y}|$ Booleans to true (resp. to false) for every leaf (resp. non-leaf) region of~$\mathcal{Y}$\\
\While{\label{line:InsertWhile}$x<\lvert \mathcal{X} \rvert$ \textbf{or}~$y<\lvert \mathcal{Y} \rvert$}{
    \If{$x<\lvert \mathcal{X} \rvert$ \textbf{and}~$y<\lvert \mathcal{Y} \rvert$ \textbf{and}~$\mathcal{X}$.map$[x] = \mathcal{Y}$.map$[y]$}{
    \tcp{Duplicate region $(x,y)$ found in~$\mathcal{X}$ and~$\mathcal{Y}$, keep (and renumber) it in $\mathcal{Z}$}
        $\text{C}_{\mathcal{X}\rightarrow\mathcal{Z}}[x] := z; \ \text{C}_{\mathcal{Y}\rightarrow\mathcal{Z}}[y] := z; \ \text{C}_{\mathcal{Z}\rightarrow\mathcal{X},\mathcal{Y}}[z] := (x,y)$\;
        $x\plusequals1; \ y\plusequals1;\ z\plusequals1$
    }
    \ElseIf{$\mathcal{Y}$.map$[y] \prec \mathcal{X}$.map$[x]$}{
        \If(\tcp*[h]{Keep (en renumber) region $y$ in~$\mathcal{Z}$}){ $InZ[y]=true$}{
            $InZ[\mathcal{Y}$.par$[y]] := true$\;
            $\text{C}_{\mathcal{Y}\rightarrow\mathcal{Z}}[y] := z; \ \text{C}_{\mathcal{Z}\rightarrow\mathcal{X},\mathcal{Y}}[z] := (-1,y)$\;
            $y\plusequals1;\ z\plusequals1$
        }
        \lElse{$y\plusequals1$ \tcp*[h]{Discard region $y$ from~$\mathcal{Z}$}}
    }
    \Else(\tcp*[h]{Keep (and renumber) region~$x$ in~$\mathcal{Z}$}){
        $\text{C}_{\mathcal{X}\rightarrow\mathcal{Z}}[x] := z; \ \text{C}_{\mathcal{Z}\rightarrow\mathcal{X},\mathcal{Y}}[z] := (x,-1)$\;
        $x\plusequals1;\ z\plusequals1$ \label{line:InsertWhileEnd}
    }
}
$\mathcal{Z} :=$ \ initialize a tree structure with $n_\mathcal{Z} = z$ nodes\;
\ForEach{\label{line:InsertFor} z \textbf{from} 0 \textbf{to} $n_\mathcal{Z}$}{
    $(x,y) := \text{C}_{\mathcal{Z}\rightarrow\mathcal{X},\mathcal{Y}}[r]$\;
    \If{$x \neq -1$}{
        $\mathcal{Z}$.map$[z] := \mathcal{X}$.map$[x]$\;
        \lIf{$[x] = \mathcal{X}$.root}{
            $\mathcal{Z}$.par$[z]:=x$
        }
        \lElse{
            $\mathcal{Z}$.par$[z] := \text{C}_{\mathcal{X}\rightarrow\mathcal{Z}}[\mathcal{X}$.par$[x]]$
        }
        \lIf{$z \geq \lvert \mathcal{X}.leaves \rvert$}{
            $\mathcal{Z}$.weight[$z - \lvert \mathcal{X}.leaves \rvert] := \mathcal{X}$.weight[$x - \lvert \mathcal{X}.leaves \rvert$]
        }
    }
    \Else{
        $\mathcal{Z}$.map$[z] := \mathcal{Y}$.map$[y]$\;
        \lIf{$y = \mathcal{Y}$.root}{
            $\mathcal{Z}$.par$[z]:=y$
        }
        \lElse{
            $\mathcal{Z}$.par$[z] := \text{C}_{\mathcal{Y}\rightarrow\mathcal{Z}}[\mathcal{Y}$.par$[y]]$
        }
        \lIf{$z \geq \lvert \mathcal{Y}.leaves \rvert$}{
            $\mathcal{Z}$.weight[$z - \lvert \mathcal{Y}.leaves \rvert] := \mathcal{Y}$.weight[$x - \lvert \mathcal{Y}.leaves \rvert$] \label{line:InsertForEnd}
        }
    }
}
\caption{{\sc Insert}}
\label{algo:insert}
\end{algorithm}

\begin{enumerate}
    \item {\em Lines \ref{line:InsertWhile}-\ref{line:InsertWhileEnd}.} Identify and renumber the regions of~$\mathcal{X}$ and~$\mathcal{Y}$ that belong to~$\mathcal{Z}$ and store the correspondences between the new number of the regions in~$\mathcal{Z}$ and the indices of the initial regions in~$\mathcal{X}$ and~$\mathcal{Y}$. It can be observed that this step is necessary since a region of~$\mathcal{Z}$ can be duplicated in both~$\mathcal{X}$ and~$\mathcal{Y}$ and that some regions of~$\mathcal{Y}$ are discarded from~$\mathcal{Z}$. In order to perform this step, the regions of~$\mathcal{X}$ and~$\mathcal{Y}$ are simultaneously browsed in increasing order for~$\prec$. The correspondences between the regions of the hierarchies are stored in three arrays:~$\text{C}_{\mathcal{X}\rightarrow\mathcal{Z}}$, $\text{C}_{\mathcal{Y}\rightarrow\mathcal{Z}}$, and $\text{C}_{\mathcal{Z}\rightarrow\mathcal{X},\mathcal{Y}}$;
    \item {\em Lines \ref{line:InsertFor}-\ref{line:InsertForEnd}.} Build the parenthood relation (\texttt{par}) of the hierarchy~$\mathcal{Z}$ using the parenthood relation of the hierarchies~$\mathcal{X}$ and~$\mathcal{Y}$ and the correspondences between the regions of the hierarchies. At the same time, we also build the attributes \texttt{map} and~\texttt{weight} associated to~$\mathcal{Z}$.
\end{enumerate}

During the first step, each region of the two hierarchies~$\mathcal{X}$ and~$\mathcal{Y}$ is considered once and processed with a limited number of constant-time instructions. Thus, the overall time complexity of Lines \ref{line:InsertWhile}-\ref{line:InsertWhileEnd} is linear with respect to the number of nodes of~$\mathcal{X}$ and~$\mathcal{Y}$. The worst-case complexity of the second step is also linear with respect to the number of nodes of~$\mathcal{X}$ and~$\mathcal{Y}$ since~$\mathcal{Z}$ contains at most all regions of each of hierarchy. Thus, the overall complexity of  Algorithm~\ref{algo:insert}  is~$O(\lvert \mathcal{X} \rvert + \lvert \mathcal{Y} \rvert)$.

\section{Conclusion}
\label{sec:conclusion}

In this article, we proposed efficient and easily implementable algorithms  for the three algebraic operations on hierarchies \emph{select}, \emph{join}, and \emph{insert}.
These algorithms rely on a particular data structure to represent local hierarchies in order to achieve linear or linearithmic time complexity while limiting the amount of information required in the main memory. Thanks to these contributions it is now possible to efficiently implement the calculus scheme proposed in~\cite{Cousty-et-al-DGMM2021} for the out-of-core computation of BPHs. In future works, we plan to study the time and memory consumption of the proposed algorithms in practice and to develop efficient algorithms to process the distribution of a BPH in order to obtain a completely out-of-core pipeline for  seeded watershed segmentation.

\subsubsection{Acknowledgements} This work  was  supported by the French ANR  grant ANR-20-CE23-0019.

%
%
%
\bibliographystyle{splncs04}
\bibliography{OOC-BPT-DGMM2022}

\end{document}